\documentclass[12pt,draft]{article}
\usepackage{amsfonts}

\def\be{\begin{equation}}
\def\ee{\end{equation}}

\begin{document}

\begin{titlepage}

\vspace{0.5cm}

\begin{center}
{\large {\bf About the self-dual Chern-Simons system and Toda field theories on the noncommutative plane}}\\
\vspace{1.5cm}
{\bf I. Cabrera-Carnero} \footnote{\tt{email:cabrera@cpd.ufmt.br}}\\
{\em Departamento de F\'{\i}sica-ICET}\\
{\em Universidade Federal de Mato Grosso}\\
{\em Av. Fernando Corr\^{e}a da Costa, $s/n^o$-Bairro Coxip\'{o}}\\
{\em 78060-900-Cuiab\'{a}-MT-Brasil}\\

\vspace{0.8cm}
\end{center}
\renewcommand{\thefootnote}{\arabic{footnote}}
\vspace{6mm}

\begin{abstract}
\noindent The relation of the noncommutative self-dual
Chern-Simons (NCSDCS) system to the noncommutative generalizations
of Toda and of affine Toda field theories is investigated more
deeply. This paper continues the programme initiated in $JHEP {\bf
0510} (2005) 071$, where it was presented how it is possible to
define Toda field theories through se\-cond order differential
equation systems starting from the NCSDCS system. Here we show
that using the connection of the NCSDCS to the noncommutative
chiral model, exact solutions of the Toda field theories can be
also constructed by means of the noncommutative extension of the
uniton method proposed in $JHEP {\bf 0408} (2004) 054$ by
Ki-Myeong Lee. Particularly some specific solutions of the nc
Liouville model are explicit constructed.

 \vspace{3cm}

\end{abstract}
\vspace{5mm}
\end{titlepage}

\newpage

\section{Introduction}

Inside the context of Noncommutative Field theories (NCFT),
noncommutative (nc) extensions of two-dimensional Integrable Field
Theories have been investigated \cite{dimakis, integrablenc,
integrablenc2, GP1, us, GP2, kimeon, lechtenfeld, zuevsky,
hamanaka, harold, me} in the last few years. Par\-ti\-cu\-lar\-ly
in \cite{me} nc extensions of Toda and affine Toda theories were
proposed. The nc extensions of Toda field theories were
constructed in \cite{me} starting from a nc zero-curvature
condition for algebra-valued potentials introduced in \cite{us}.
Expressing the gauge potentials in a particular way this condition
can be reduced to the nc Leznov-Saveliev equation, which as shown
in \cite{me} can be regarded as the equation of motion of a
constrained nc Wess-Zumino-Novikov-Witten model ($WZNW_{\star}$).
In this sense the extensions of abelian and abelian affine Toda
theories presented in \cite{me} have the advantage of possessing
an infinite number of conserved charges and of being represented
by second order differential equations for $N$ fields. The
corresponding action principles were also presented in \cite{me}.

On the other hand in \cite{kimeon} was considered the nc extension
 to the nc plane of the Dunne-Jackiw-Pi-Trugenber (DJPT) model \cite{dunne1} of a
$U(N)$ Chern-Simons gauge theory coupled to a nonrelativistic
complex adjoint matter field. The lowest energy solutions of this
model satisfy a nc extension of the self-dual Chern-Simons
equations. Through a proposed ansatz, nc generalizations of $U(N)$
Toda and affine Toda theories were constructed from these nc
self-dual equations. The generalizations of Toda theories proposed
were expressed as systems of first order differential equations
for $2N-1$ fields which could not be reduced to coupled second
order equations in general. The advantage of defining the Toda
field theories in this way was the possibility of constructing
exact solutions since the self-dual equations for Chern-Simons
solitons on nc space can be related to the equation of the $U(N)$
nc chiral model, which apparently can be also solved by the
Uhlenbeck's uniton method \cite{uhlenbeck} as was suggested in
\cite{kimeon}. In \cite{me} was shown that the NCSDSC system can
be reduced to the nc Leznov-Saveliev equations using a different
ansatz and in this way obtain the Toda field theories as second
order differential equations. In this paper we would like to
present the relation between the NCSDCS system and the nc Toda
field theories in a more detailed way, essentially in connection
with the construction of exact solutions that as we will see is
still possible. In this sense this paper complements the results
presented in the last section of \cite{me}.

This paper is organized as follows. In the first section we review
the derivation of the nc extensions of abelian and abelian affine
Toda field theories presented in \cite{me}. In Section $3$ we
present how the NCSDCS system can be transformed into the nc
Leznov-Saveliev equations from where the nc extensions of Toda
field theories were constructed in \cite{me}. In this section we
also show how the system of first order differential equations for
$2N-1$ fields considered \cite{kimeon} as the nc extension of Toda
field theories can be reduced to our nc extension of Toda
theories, i.e. a system of second order differential equations for
$N$ fields. In section $4$ the relation of the NCSDCS to the nc
principal chiral model \cite{kimeon} is reviewed. The nc extension
of the uniton method proposed in \cite{kimeon} and the explicit
construction of some solutions of our nc extension of Liouville
model through this method is as well exposed in this section. The
last section provides the conclusions.

\section{Toda theories from $WZNW_{\star}$}

It is well known that Toda theories connected with finite simple
Lie algebras, on the ordinary commutative case, can be regarded as
constrained Wess-Zumino-Novikov-Witten ($WZNW$) models
\cite{balog}\footnote {Affine Toda theories can be as well
regarded as  contrained two-loop $WZNW$ models \cite{aratyn}.}. By
placing certain constraints on the chiral currents, the
G-invariant $WZNW$ model reduces to the appropriate Toda theory.
Specifically, the abelian Toda theories are connected with abelian
embeddings $G_0\subset G$. In \cite{me} we constructed nc
extensions of abelian and abelian affine Toda theories applying
this procedure to the nc extension of the $WZNW$ model
($WZNW_{\star}$). Here we will briefly review our results.

As usually NCFT \cite{reviews} can be constructed from a given
field theory by replacing the product of fields by an associative
$\star$-product. Considering that the noncommutative parameter
$\theta^{\mu \nu}$ is a constant antisym\-me\-tric tensor, the
deformed product of functions is expressed trough the Moyal
product \cite{moyal} : \be
  \phi_1(x)\phi_2(x) \to
\phi_1(x)\star\phi_2(x)=e^{\frac{i}{2}\theta^{\mu\nu}\partial_\mu^{x_1}
\partial_\nu^{x_2}}\phi_1(x_1)\phi_2(x_2)|_{x_1=x_2=x}.
\ee
In the following we will refer to functions of operators in
the noncommutative deformation by a $\star$ sub-index, for example
$e^{\phi}_\star=\sum_{n=1}^{\infty}\frac{1}{n!}\phi_\star^{n}$
(the n-times star-product of $\phi$ is understood).

 Consider now the nc generalization of the $WZNW$ model introduced
in \cite{wznwnc}
\begin{eqnarray}
S_{WZNW_{\star}}&=&-\frac{k}{4\pi}\int_{\Sigma} d^2z Tr (
g^{-1}\star
\partial{g} \star g^{-1} \star \bar{\partial}g)+ \nonumber \\
&&\frac{k}{24\pi}\int_{B} d^3 x \epsilon_{ijk}(g^{-1}\star
\partial_i g \star g^{-1}\star
\partial_j g \star g^{-1}\star \partial_k g ). \label{wznwnc}
\end{eqnarray}
Here B is a three-dimensional manifold whose boundary $\partial
B=\Sigma $. We are considering that
 $z=t+x$, $\bar{z}=t-x$ and $\partial=\frac{1}{2}(\frac{\partial}{\partial t}+\frac{\partial}{\partial
 x})$, $\bar{\partial}=\frac{1}{2}(\frac{\partial}{\partial t}-\frac{\partial}{\partial
 x})$. The coordinates $z, \bar{z}$ or
equivalently $x,t$ are noncommutative, but the extended coordinate
$y$ on the manifold $B$ remains commutative, i.e.
$[z,\bar{z}]=\theta$, $[y,z]=[y,\bar{z}]=0$. The Euler-Lagrange
equations of motion corresponding to (\ref{wznwnc}) are
\begin{eqnarray}
 \bar{\partial}J=\partial \bar{J}=0, \label{eqnwznwnc}
 \end{eqnarray}
where $J$ and $\bar{J}$ represent the conserved chiral currents
\begin{eqnarray}
J=g^{-1}\star \partial g,  \quad \quad \quad
\bar{J}=-\bar{\partial}g \star g^{-1}. \label{chiralcurrents}
\end{eqnarray}
The fields $\alpha_a$ parameterize the group element $g \in G$
through $ g=e_{\star}^{\alpha_a T_a}$, where $T_a$ are the
generators of the corresponding algebra ${\cal G}$. As it was our
interest to define the theories inside a $G_0$ subgroup of $G$,
the unwanted degrees of freedom that corresponds to the tangent
space $G/G_0$ were eliminated implementing constraints upon
specific components of the currents $J, \bar{J}$:
\begin{eqnarray}
J_{constr}=j+\epsilon_-, \quad \quad \quad
\bar{J}_{constr}=\bar{j}+\epsilon_{+}, \label{HRconstraints}
\end{eqnarray}
where $\epsilon_{\pm}$  are constant elements of grade $\pm 1$
with respect to a grading operator $Q$, i.e.
$[Q,\epsilon_{\pm}]=\pm \epsilon_{\pm}$ defined in the algebra
$\cal{G}$ and $j,\bar{j}$ contain generators of grade zero and
positive, and zero and negative respectively. The grading operator
$Q$ decomposes the algebra ${\cal G}$ in $\mathbb{Z}$-graded
subspaces,
 \begin{eqnarray}
[Q,{\cal G}_i]=i{\cal G}_i, \quad \quad  [{\cal G}_i,{\cal G}_j]
\in {\cal G}_{i+j}.
 \end{eqnarray}
This means that the algebra ${\cal G}$ can be represented as the
direct sum $ {\cal G}=\bigoplus_i {\cal G}_i$ and that the
subspaces ${\cal G}_0,{\cal G}_>,{\cal G}_{<}$ are subalgebras of
${\cal G}$, composed of the Cartan and of the positive/negative
steps generators respectively.
 The algebra can then
be written using the triangular decomposition $
 {\cal G}={\cal G}_<\bigoplus {\cal G}_0 \bigoplus {\cal G}_> $. Denote the subgroup elements obtained through the
$\star$-exponentiation of the ge\-ne\-ra\-tors of the
corresponding subalgebras as $N=e^{{\cal G}_<}_{\star}$, $
B=e^{{\cal G}_0}_{\star}$, $M=e^{{\cal G}_>}_{\star}$. Then
proposing a nc Gauss-like decomposition, an element $g$ of the nc
group $G$ can be expressed as
 \be
 g=N \star B \star M. \label{Gaussnc}
 \ee
The reduced model is obtained then introducing (\ref{Gaussnc}) in
(\ref{chiralcurrents}) and giving the constant elements
$\epsilon_{\pm}$ of grade $\pm 1$ responsible for
  cons\-training the currents in a general manner. As result of the reduction
process, the degrees of freedom in $M,N$ are eliminated and the
equations of motion of the constrained model are natural nc
extensions of the Leznov-Saveliev equations of motion
\cite{LezSav}, namely
\begin{eqnarray}
\bar{\partial}(B^{-1}\star\partial
B)+[\epsilon_-,B^{-1}\star\epsilon_+ B]_{\star}=0,
\nonumber \\
\partial(\bar{\partial}B \star B^{-1})-[\epsilon_+,B \epsilon_-\star
B^{-1}]_{\star}=0. \label{Lez-Savnc}
\end{eqnarray}
As shown in \cite{us}, the equations of motion (\ref{Lez-Savnc})
can be expressed as a generalized $\star$-zero-curvature condition
 \be
 \bar{\partial}A-\partial
\bar{A}+[A,\bar{A}]_{\star}=0, \label{starzc} \ee
 taking the potentials as
$A=-B \epsilon_-\star B^{-1}$ and $
\bar{A}=\epsilon_++\bar{\partial}B \star B^{-1}$. This condition
(\ref{starzc}) implies the exis\-ten\-ce of an in\-fi\-ni\-te
amount of conserved charges \cite{us}. For this reason in order to
preserve the original integrability properties of the
two-dimensional models (\ref{Lez-Savnc}) can be a reasonable
starting point in order to construct nc analogs of Toda models.

The eqs. (\ref{Lez-Savnc}) are the Euler-Lagrange equations of
motion of the action
 \be
S=S_{WZNW_{\star}}(B)+\frac{k}{2\pi}\int d^2z Tr (\epsilon_+\star
B \epsilon_-\star B^{-1}). \label{effectiveactionnc}
 \ee

From (\ref{Lez-Savnc}) we constructed in \cite{me} nc analogs of
the $GL(n,\mathbb{R})$ abelian Toda theories taking the gradation
operator as $
 Q=\sum_{i=1}^{n-1}\frac{2\lambda_i
\cdot H}{\alpha_i^2}$, where $H$ re\-pre\-sents the Cartan
subalgebra, $\alpha_i$ is the $i^{th}$ simple root and $\lambda_i$
is the $i^{th}$ fundamental weight that satisfies $
\frac{2\lambda_i \cdot \alpha_j}{\alpha_i^2}=\delta_{ij}$. In this
way it defines the subalgebra of grade zero as ${\cal
G}_0=U(1)^{n}=\{I,h_i,\, \,\,i=1 \dots n-1\}$, where the Cartan
generators are defined in the Chevalley basis as $h_i=\frac{2
\alpha_i \cdot H}{\alpha_i^2}$. The zero grade group element $B$
is then expressed through the $\star$-exponentiation of the
generators of the zero grade subalgebra ${\cal G}_0$, i.e. the
${\cal SL}(n)$ Cartan subalgebra plus the identity generator,
\begin{eqnarray}
B=e^{\Sigma_{i=1}^{n-1}\varphi_i h_i + \varphi_0 I}_{\star}.
\label{Bnctoda}
\end{eqnarray}
With the constant generators of grade $\pm 1$ as
\begin{eqnarray}
\epsilon_{\pm}= \Sigma_{i=1}^{n-1}\mu_i E_{\pm \alpha_i},
\label{eps}
\end{eqnarray}
where $E_{\pm \alpha_i}$ are the step generators associated to the
positive/negative simple roots of the algebra and $\mu_i$ are
constant parameters, the equations of motion that define the nc
extension of abelian Toda models are
\begin{eqnarray}
\partial(\bar{\partial}( e_{\star}^{\phi_{k}})\star
e_{\star}^{-\phi_{k}})&=& \mu_{k}^2 e_{\star}^{\phi_{k+1}}\star
e_{\star}^{-\phi_{k}}-\mu_{k-1}^2e_{\star}^{\phi_{k}}\star
e_{\star}^{-\phi_{k-1}},
 \label{todaeqnmotion}
\end{eqnarray}
a system of $n$-coupled equations ($k=1\dots n$) where we have
changed va\-ria\-bles to:
\begin{eqnarray}
\varphi_1+\varphi_0&=&\phi_1, \nonumber \\
-\varphi_k+\varphi_{k+1}+\varphi_0&=&\phi_{k+1}, \, \, \, \,
\mathrm{for} \, \, k=1 \, \, \mathrm{to} \, \, n-2,
\label{newvariables}
\\
-\varphi_{n-1}+\varphi_0&=&\phi_n. \nonumber
\end{eqnarray}
 Notice that for the first and last equation $\mu_0=\mu_n=0$ and
$\phi_0=\phi_{n+1}=0$. A particular example of these field
theories is the nc extension of the Liouville model:
\begin{eqnarray}
\partial( \bar{\partial}(e_{\star}^{\phi_+})\star
e^{-\phi_+}_{\star})
=\mu^2 e^{\phi_-}_{\star}\star e^{-\phi_+}_{\star}, \nonumber \\
\partial(\bar{\partial}(e_{\star}^{\phi_-})\star
e^{-\phi_-}_{\star})=-\mu^2 e^{\phi_-}_{\star}\star
e^{-\phi_+}_{\star}, \label{eqnliouvillenc2}
\end{eqnarray}
with $\phi_1=\phi_+$ and $\phi_2=\phi_-$. This system in the
commutative limit will lead to a decoupled model of two fields: a
free field and the usual Liouville field, i. e.
\begin{eqnarray}
\partial \bar{\partial} \varphi_0=0 \quad \mathrm{and} \quad \partial
\bar{\partial} \varphi_1=\mu^2 e^{-2\varphi_1},
\label{liouvillecommutative}
\end{eqnarray}
where $\varphi_1=\phi_+-\phi_-$ and $\varphi_0=\phi_++\phi_-$. The
action, whose Euler-Lagrange equations of motion leads to
(\ref{todaeqnmotion}), can be obtained from
(\ref{effectiveactionnc}) with (\ref{Bnctoda}) and (\ref{eps}). It
reads
\begin{eqnarray}
S(\phi_1,\dots,\phi_n)=
\sum_{k=1}^{n}S_{WZNW_{\star}}(e_{\star}^{\phi_k})+\frac{k}{2\pi}\int
d^2z \sum_{k=1}^{n-1}\mu^2_k(e_{\star}^{\phi_{k+1}}\star
e_{\star}^{-\phi_{k}}). \label{todancaction2}
\end{eqnarray}

In the same way the nc extensions of
$\widetilde{GL}(n,\mathbb{R})$\footnote{We refer to the loop
algebra, see \cite{me} for more details.} abelian affine Toda
theories can be constructed taking the gradation operator as $
 Q=\sum_{i=1}^{n-1}\frac{2\lambda_i \cdot
H^{(0)}}{\alpha_i^2}+n d$, where $d$ is the derivation generator
whose coefficient is chosen such that this gradation ensures that
the zero grade subspace ${\cal G}_0$ coincides with its
counterpart on the corresponding Lie algebra ${\cal
SL}(n,\mathbb{R})$, apart from the generator $d$. The major
di\-ffe\-ren\-ce with the finite case is in the constant
generators of grade $\pm 1$ that include extra affine generators,
say
\begin{eqnarray}
\epsilon_{\pm}=\sum_{i=1}^{n-1}\mu_iE_{\pm \alpha_i}^{(0)}+m_0
E^{(\pm1)}_{\mp\psi}, \label{epsaffine}
\end{eqnarray}
here $\psi$ is the highest root of ${\cal G}={\cal
SL}(n,\mathbb{R})$ and  $m_0$, $\mu_i$ with $i=1\dots n-1$ are
constant parameters. Using again the nc extension of the
Leznov-Saveliev equations of motion (\ref{Lez-Savnc}) the nc
analogs of abelian affine $\widetilde{GL}(n,\mathbb{R})$ Toda
theories are obtained
\begin{eqnarray}
\partial(\bar{\partial}( e_{\star}^{\phi_{k}})\star
e_{\star}^{-\phi_{k}})= \mu_{k}^2e_{\star}^{\phi_{k+1}}\star
e_{\star}^{-\phi_{k}} &-&\mu_{k-1}^2 e_{\star}^{\phi_{k}}\star
e_{\star}^{-\phi_{k-1}}- \label{todaeqnaffinenc} \\
&+&m^2_0(\delta_{n,k}-\delta_{1,k}) e_{\star}^{\phi_1}\star
e_{\star}^{-\phi_n}. \nonumber
\end{eqnarray}
Note that in the previous expression $\mu_0=\mu_n=0$ and $k=1\dots
n$. The action from where the nc $\widetilde{GL}(n)$ affine
equations (\ref{todaeqnaffinenc}) can be derived reads
\begin{eqnarray}
S(\phi_1,\dots,\phi_n)=&&\sum_{k=1}^{n}S_{WZNW_{\star}}(e_{\star}^{\phi_k})+ \label{todancactionaffine} \\
&& +\frac{k}{2\pi}\int d^2z \left(\sum_{k=1}^{n-1}\mu^2_k
e_{\star}^{\phi_{k+1}} \star
e_{\star}^{-\phi_{k}}+m^2_0e_{\star}^{\phi_1}\star
e_{\star}^{-\phi_n}\right). \nonumber
\end{eqnarray}
Among these theories it is found the nc extension of the
sine-Gordon model:
\begin{eqnarray}
\partial(\bar{\partial}(e^{\phi_+}_{\star})\star
e^{-\phi_+}_{\star}) &=&\mu^2(e_{\star}^{\phi_-}\star
e_{\star}^{-\phi_+}-e_{\star}^{\phi_+}\star e_{\star}^{-\phi_-}), \nonumber \\
\partial(\bar{\partial}(e^{\phi_-}_{\star})\star
e^{-\phi_-}_{\star})&=&\mu^2(-e_{\star}^{\phi_-}\star
e_{\star}^{-\phi_+}+e_{\star}^{\phi_+}\star e_{\star}^{-\phi_-}).
\label{eqnsinegordonnc2}
\end{eqnarray}

Let us remark that by {\it abelian} we refer to a property of the
original ordinary commutative theory. On the noncommutative
scenario it happens that the zero grade subgroup $G_0$ despite it
is spanned by the generators of the Cartan subalgebra, turns out
to be nonabelian, i.e, if $g_1,g_2$ are two elements of the zero
grade subgroup $G_0$ then $g_1 \star g_2 \neq g_2 \star g_1$.

 As a resume we constructed in \cite{me} nc extensions of $GL(n,\mathbb{R})$
abelian and $\widetilde{GL}(n,\mathbb{R})$ abelian affine Toda
theories as systems of second order differential equations for $n$
fields. The nc models differentiate from the commutative case by
the presence of an extra field which will not decouple in the
equations of motion. The presence of this extra field is due to
the introduction of the the identity generator in the Cartan
subalgebra since the algebra ${\cal SL}(n)$ has been extended to
${\cal GL}(n)$.

\section{NC Self-Dual Chern-Simons system}

The Chern-Simons theories in the ordinary commutative space have
played a central role in the understanding of different phenomena
in planar physics. Recently they have been extended to
noncommutative spaces (see for example the refs.
\cite{chernsimons,poly}), where apparently they have proven to be
also useful for the description of different phenomena, as for
example, the Quantum Hall Effect \cite{susskind}. Inside this
context the nc self-dual Chern-Simons system:
 \begin{eqnarray}
\bar{D} \Psi&=&0, \nonumber \\
\bar{\partial}
A-\partial\bar{A}+[\bar{A},A]_{\star}&=&\frac{1}{k}[\Psi^{\dag},
\Psi]_{\star}, \label{selfdualchsimons}
\end{eqnarray}
with $D=\partial+[A, \,]_{\star}$ and
$\bar{D}=\bar{\partial}+[\bar{A}, \,]_{\star}$ the covariant
derivatives, have been considered in different works
\cite{selfdualchernsimons}. Particularly in \cite{kimeon}, was
considered the nc extension of the Dunne-Jackiw-Pi-Trugenber
(DJPT) \cite{dunne1}(see also refs. \cite{dunne2,dunne3}) model of
a $U(N)$ Chern-Simons gauge theory coupled to a
nonre\-la\-ti\-vis\-tic complex bosonic matter field on the
adjoint representation. The lowest energy soliton solutions of
this model satisfy (\ref{selfdualchsimons}) and they are related
to the exact solutions of the $U(N)$ nc chiral model. Through a
proposed ansatz nc generalizations of $U(N)$ Toda and affine Toda
theories were constructed \cite{kimeon}. Although in the
commutative case this procedure will lead to the well known second
order differential equations of the Conformal Toda or affine Toda
theories \cite{dunne1}, in the noncommutative scenario the
generalization of Toda theories proposed in \cite{kimeon} were
expressed as systems of first order differential equations for
$2N-1$ fields which could not be reduced to coupled second order
equations in ge\-ne\-ral.

In this section we would like to relate our nc extensions of the
abelian and abelian affine Toda models (\ref{todaeqnmotion}),
(\ref{todaeqnaffinenc}) and the proposal presented in
\cite{kimeon} for these Toda models. We will see in the following
that the nc Leznov-Saveliev equations (\ref{Lez-Savnc}) can be
obtained from the nc Chern-Simons self-dual soliton equations
 (\ref{selfdualchsimons}) expressing the gauge potentials $A, \bar{A}$ and the matter field
 $\Psi$ in a particular way. This
gives the possibility of obtaining nc extensions of Toda and
affine Toda models as second order differential equations.  We
will also see that using the equivalence of the nc self-dual
equations to the nc chiral model equation it is possible to
construct exact solutions of the nc Toda models
(\ref{todaeqnmotion}).

From now on we will use the operator formalism. This language is
sometimes more convenient since the nonlocality of the star
product renders explicitly calculations quite complicated.

 The complex coordinates $z=t+ix$
and $\bar{z}=t-ix$ satisfy $[z, \bar{z}]=\theta$. This suggests
that we can represent $z, \bar{z}$ as creation and annihilation
operators $a=z$, $a^{\dag}=\bar{z}$. These operators will act on
the harmonic-oscillator Fock space ${\cal H}$ with an orthonormal
basis $|n>=\frac{(a^{\dag})^n}{\sqrt{n!}}|0>,$ for $n=0,1,2,
\dots$ such that the vacuum is defined as $a|0>=0$. Further
\begin{eqnarray}
 a|n>=\sqrt{\theta n}|n-1>, \quad a^{\dag}|n>=\sqrt{\theta(n+1)}|n+1>,
\end{eqnarray}
and the number operator is defined as
$a^{\dag}a|n>=:\mathbb{N}|n>=\theta n|n>$. Any function
$f(t,z,\bar{z})$ on the nc space can be related to an operator
valued function $\hat{f}(t)\equiv F(t,a, a^{\dag})$ acting on
${\cal H}$ by use of the Weyl transform \cite{harvey}.
 In the operator formalism any field on the nc space
becomes an operator on the Hilbert space and the
 derivatives as well can be represented by operators
\begin{eqnarray}
\partial=\frac{1}{2}(\partial_t-i\partial_x)=\partial_z=-\frac{1}{\theta}[a^{\dag},
\quad], \quad
\bar{\partial}=\frac{1}{2}(\partial_t+i\partial_x)=\partial_{\bar{z}}=\frac{1}{\theta}[a,
\quad]. \label{derivatives}
\end{eqnarray}
Let us consider the self-dual Chern-Simons system in the operator
formalism:
\begin{eqnarray}
\bar{D} \hat{\Psi}&=&0, \nonumber \\
\frac{1}{\theta}
[a,\hat{A}]+\frac{1}{\theta}[a^{\dag},\hat{\bar{A}}]+[\hat{\bar{A}},\hat{A}]&=&\frac{1}{k}[\hat{\Psi}^{\dag},
\hat{\Psi}]. \label{selfdualchsimonsop}
\end{eqnarray}

In the following we will see how the operator version of the nc
Leznov-Saveliev equations (\ref{Lez-Savnc}) can be also obtained
from (\ref{selfdualchsimonsop}). For this purpose let us consider
that the gauge fields are expressed as
\begin{eqnarray}
\hat{A} =-\frac{1}{\theta}\hat{G}^{-1}[a^{\dag},\hat{G}], \quad
\hat{\bar{A}}=-\hat{A}^{\dag}, \label{A}
\end{eqnarray}
where $\hat{G}$ is an element of the complexification of the gauge
group $G$. Suppose we can decompose $\hat{G}$ as
 \be
\hat{G}=\hat{H}\hat{U},
 \ee
where $\hat{H}$ is hermitian and $\hat{U}$ is unitary. The field
strength is then expressed as
 \begin{eqnarray}
 \hat{F}_{+-}=\frac{1}{\theta}[a, \hat{A}]+\frac{1}{\theta}[a^{\dag},
 \hat{\bar{A}}]+[\hat{\bar{A}},\hat{A}]=-\frac{1}{\theta^2}U^{-1}H [a,H^{-2}[a^{\dag},H^2]]H^{-1}U
\end{eqnarray}
and the solution of the self-duality equation $\bar{D}
\hat{\Psi}=0$ is trivially: \be \hat{\Psi}=\sqrt{k} \hat{G}^{-1}
\hat{\Psi}_0(a) \hat{G}, \label{psi} \ee for any
$\hat{\Psi}_0(a)$. Inserting this solution in the other
self-duality equation (\ref{selfdualchsimons}) yields the equation
for $\hat{H}$:
 \be
-\frac{1}{\theta^2}[a,\hat{H}^{-2}[a^{\dag},\hat{H}^2]]=\hat{\Psi}_0^{\dag}
\hat{H}^{-2} \hat{\Psi}_0 \hat{H}^{2}-\hat{H}^{-2}\hat{\Psi}_0
\hat{H}^{2}\hat{\Psi}_0^{\dag}. \ee What is a second order
differential equation for the fields that parameterized
$\hat{H}^2=\hat{B}$, i. e. the elements of the zero-grade
subgroup. Considering that $\hat{\Psi}_0=\epsilon_+$, i.e. the
generator of grade $\pm 1$ which satisfy
$\epsilon_-^{\dag}=\epsilon_+$, the previous equation is written
as
\begin{eqnarray}
-\frac{1}{\theta^2}[a,\hat{B}^{-1}[a^{\dag},\hat{B}]]-
[\epsilon_-,\hat{B}^{-1}\epsilon_+ \hat{B}]=0, \label{Lez-Savnco}
\end{eqnarray}
which could be the operator version of the nc Leznov-Saveliev
equation (\ref{Lez-Savnc}), as can be tested using the Weyl-Moyal
map \cite{harvey} unless the minus sign in front of the second
term. This procedure is a nc extension of an alternative way for
obtaining the Toda models from the Chern-Simons self-dual
equations presented in \cite{dunne1} and it allows to define the
Toda models as second order differential equations, as we will see
in the following.

\subsection{NC self-dual Chern-Simons and Toda field theories}

The NCSDCS system (\ref{selfdualchsimonsop}), as was shown in
\cite{me}, can be obtained from the nc self-dual Yang-Mills
equations in four dimensions through a dimensional reduction
process. In this section we will define the Toda field theories
(\ref{todaeqnmotion}) starting from this system. Hence this is
another example where the Ward conjecture \cite{ward} apparently
also works on the nc scenario. Others nc extensions of
two-dimensional integrable models have been also derived from the
$D=4$ nc self-dual Yang-Mills equations \cite{integrablenc2,
hamanaka}.

In \cite{kimeon} was proposed for $U(N)$ the ansatz
\begin{eqnarray}
\hat{A}=\mathrm{diag}(\hat{E}_1,\hat{E}_2,...,\hat{E}_N), \quad
\hat{\Psi}_{ij}=\delta_{i,j-1}\hat{h}_i \quad i=1,...,N-1,
\label{kime}
\end{eqnarray}
which after introducing in (\ref{selfdualchsimonsop}) leads to a
system of coupled first order equations for the fields $\hat{E}_i$
with $i=1,\dots ,N$ and for the fields $\hat{h}_i$ with $i=1,\dots
,N-1$:
\begin{eqnarray}
&&\frac{1}{\theta}[a,\hat{h}_i]-\hat{E}_i^{\dag}\hat{h}_i+\hat{h}_{i}\hat{E}_{i+1}^{\dag}=
 0, \quad \mathrm{for} \quad
i=1,2, \dots N-1, \nonumber \\
&-&\frac{1}{\theta}\left[a^{\dag}, \hat{E}_1^{\dag}\right]+\frac{1}{\theta}[a, \hat{E}_1]+[\hat{E}_1^{\dag}, \hat{E}_1]= -\hat{h}_1 \hat{h}_1^{\dag}, \nonumber \\
&-&\frac{1}{\theta}\left[a^{\dag},
\hat{E}_i^{\dag}\right]+\frac{1}{\theta}[a,
\hat{E}_i]+[\hat{E}_i^{\dag},\hat{E}_i]=\hat{h}_{i-1}^{\dag}
\hat{h}_{i-1}-\hat{h}_i \hat{h}_i^{\dag}, \, \mathrm{for} \,
i=2,..,N-1, \nonumber \\
&-&\frac{1}{\theta} \left[a^{\dag},
\hat{E}_N^{\dag}\right]+\frac{1}{\theta}[a,
\hat{E}_N]+[\hat{E}_N^{\dag},
 \hat{E}_N]=\hat{h}_{N-1}^{\dag} \hat{h}_{N-1}, \label{kitoda}
\end{eqnarray}
where we have considered $k=1$. In the ordinary commutative case
the corres\-pon\-ding system is reduced to a system of second
order differential equations for the fields $\hat{h}_i$ with $i=1
\dots N-1$ only. Since the first set of equations in the above
system can not be solved for the fields $\hat{E}_i$ with $i=1\dots
N$, the system (\ref{kitoda}) can not be reduced to second order
differential equations. Looking at (\ref{A}) we see that $\hat{A}$
is expressed in terms of first order derivatives
\begin{eqnarray}
\hat{A}=-\frac{1}{\theta}\hat{U}^{-1}\hat{H}^{-1}[a^{\dag},\hat{H}]\hat{U}-\frac{1}{\theta}\hat{U}^{-1}[a^{\dag},\hat{U}]
\label{AA}.
\end{eqnarray}
For $U(N)$ we can take the constant generators as
$\epsilon_{\pm}=\sum_{i=1}^{n-1}E_{\pm \alpha_i}$. As $\hat{B}$ is
an element of the zero grade subspace it can be represented by a
diagonal matrix \be
\hat{B}=\mathrm{diag}(\hat{g}_1,\hat{g}_2,\dots, \hat{g}_N). \ee
 In order to map the systems (\ref{todaeqnmotion}) and (\ref{kitoda}) we will consider that the unitary matrix $U$ is the identity
 matrix, i.e. $U=I$. Now it is possible to choose
\be \hat{\Psi}=\hat{H} \hat{\Psi}_0(a) \hat{H}^{-1}, \label{psi1}
\ee what will lead to \be
-\frac{1}{\theta^2}[a,\hat{H}^{-2}[a^{\dag},\hat{H}^2]]=-\hat{\Psi}_0
\hat{H}^{-2} \hat{\Psi}_0^{\dag}
\hat{H}^{2}+\hat{H}^{-2}\hat{\Psi}_0^{\dag}
\hat{H}^{2}\hat{\Psi}_0, \ee that can be exactly transformed on
the operator version of the nc Leznov-Saveliev equation
\cite{LezSav},
\begin{eqnarray}
-\frac{1}{\theta^2}[a,\hat{B}^{-1}[a^{\dag},\hat{B}]]+
[\epsilon_-,\hat{B}^{-1}\epsilon_+ \hat{B}]=0, \label{Lez-Savncoo}
\end{eqnarray}
considering $\hat{\Psi}_0=\epsilon_-$. Now we are in position of
mapping the models. Thus,
\begin{eqnarray}
\hat{\Psi}_{ij}=\delta_{i-1,j}\hat{g}_{i}^{1/2}\hat{g}^{-1/2}_{i-1}.
\end{eqnarray}
and the relations
\begin{eqnarray}
\hat{h}_i^{\dag}=\hat{g}_{i+1}^{1/2}\hat{g}^{-1/2}_{i}, \quad
&& \mathrm{for} \quad i=1\dots N-1, \nonumber \\
\hat{E}_i=-\frac{1}{\theta}\hat{g}^{-1/2}_{i}[a^{\dag},\hat{g}^{1/2}_{i}],
\quad && \mathrm{for} \quad i=1\dots N, \label{relations1}
\end{eqnarray}
are obtained. If we introduce the above relations on the system
(\ref{kitoda}) it is not difficult to see that it reduces to
\begin{eqnarray}
-\frac{1}{\theta^2}\left[a, \hat{g}_1^{-1}[a^{\dag},\hat{g}_1]\right]&=&\hat{g}_1^{-1}\hat{g}_{2}, \nonumber \\
 -\frac{1}{\theta^2}\left[a, g_i^{-1}[a^{\dag},\hat{g}_i]\right]&=&\hat{g}_i^{-1}\hat{g}_{i+1}-\hat{g}_{i-1}^{-1}\hat{g}_i, \quad
\mathrm{for} \quad i=2\dots N-1, \nonumber \\
-\frac{1}{\theta^2}\left[a,\hat{g}_N^{-1}[a^{\dag},\hat{g}_N]\right]&=&-\hat{g}_{N-1}^{-1}\hat{g}_{N},
\label{Todaop}
\end{eqnarray}
which is the operator version of the Toda model
(\ref{todaeqnmotion}). The first equations in (\ref{kitoda}) are
trivially satisfied since $\hat{\Psi}$ was chosen as a solution of
these equations. The simplest example of the $U(N)$ Toda field
theories is the nc Liouville model which corresponds to $N=2$. In
this case we can write \be \hat{B}=\left(\begin{array}{cc}
     \hat{g}_+ & 0 \\
         0 & \hat{g}_-
\end{array} \right), \label{Bo}
\ee and the constant generators $\epsilon_{\pm}=E_{\pm \alpha}$.
The equations of motion from (\ref{Todaop}) or equivalently from
(\ref{Lez-Savnco}) for this model will be
\begin{eqnarray}
\frac{1}{\theta^2}\left[a, \hat{g}_+^{-1}
[a^{\dag},\hat{g}_+]\right] =-\hat{g}_+^{-1}\hat{g}_-, \quad
\frac{1}{\theta^2}\left[a, \hat{g}_-^{-1}
[a^{\dag},\hat{g}_-]\right]=\hat{g}_+^{-1}\hat{g}_-.
\label{eqnliouvillenc0}
\end{eqnarray}
In the same way it is possible to consider the affine models. In
\cite{kimeon} the affine ansatz considered  was
\begin{eqnarray}
\hat{A}&=&\mathrm{diag}(\hat{E}_1,\hat{E}_2,...\hat{E}_N), \\
\hat{\Psi}_{ij}&=&\delta_{i,j-1}\hat{h}_i, \quad i=1...N-1,
 \quad \mathrm{except}
\quad \mathrm{for} \quad \hat{\Psi}_{N1}=\hat{h}_N \nonumber
\label{kimeaffine}.
\end{eqnarray}
Here again we can established relations analogs to
(\ref{relations1}) using (\ref{AA}) and (\ref{psi1}), but now
remembering that $\epsilon_{\pm}=\sum_{i=1}^{n-1}E_{\pm
\alpha_i}^{(0)}+ E^{(\pm1)}_{\mp\psi}$. The relations obtained are
essentially the relations (\ref{relations1}), except the component
$(\hat{\Psi}^{\dag})_{1N}=\hat{h}_{N1}^{\dag}=\hat{g}_1^{\frac{1}{2}}\hat{g}_{N}^{-\frac{1}{2}}$
coming from the extra affine generator.

\section{The nc chiral model and the solutions}

In the commutative scenario the equivalence of the self-dual
Chern-Simons equations and the chiral model equation is
significative in the sense that all the solutions of the later
have been classified \cite{uhlenbeck}. In \cite{kimeon} was
investigated the extension of this relation to a nc space-time. As
we will employ the uniton method for the construction of exact
solutions of the Toda field theories (\ref{Todaop}) we will
present the relation among the nc chiral model and the nc
self-dual Chern-Simons system (\ref{selfdualchsimonsop}) in the
following. Hence, let us consider the new gauge connections,
\begin{eqnarray}
\hat{{\cal A}}\equiv \hat{A}-\sqrt{\frac{1}{k}}\hat{\Psi}, \quad
\hat{\bar{{\cal A}}}\equiv
\hat{\bar{A}}+\sqrt{\frac{1}{k}}\hat{\Psi}^{\dag},
\label{gaugepot}
\end{eqnarray}
which satisfy a zero-curvature condition
\begin{eqnarray}
\frac{1}{\theta}[a, \hat{{\cal
A}}]+\frac{1}{\theta}[a^{\dag},\hat{\bar{{\cal
A}}}]+[\hat{\bar{{\cal A}}}, \hat{{\cal A} }]=0.
\end{eqnarray}
This means that we can write $\hat{{\cal A}}, \hat{\bar{\cal A}}$
as pure gauge
\begin{eqnarray}
\hat{{\cal A}}=-\frac{1}{\theta}\hat{g}^{-1}[a^{\dag},\hat{g}],
\quad \hat{\bar{{\cal
A}}}=\frac{1}{\theta}\hat{g}^{-1}[a,\hat{g}],
\end{eqnarray}
for $\hat{g}$ in some $U(N)$. Defining
$\hat{\chi}=\sqrt{\frac{1}{k}}\hat{g}\hat{\Psi} \hat{g}^{-1}$ the
nc self-dual Chern-Simons system (\ref{selfdualchsimons}) can be
converted into a single equation
 \be
 \frac{1}{\theta}[a,\hat{\chi}]=[\hat{\chi}^{\dag}, \hat{\chi}], \label{chi} \ee
 since
 \begin{eqnarray}
\bar{D}\Psi=\sqrt{k}g^{-1}\left(\frac{1}{\theta}[a,\chi]-[\chi^{\dag},
\chi] \right)
\end{eqnarray}
and
\begin{eqnarray}
 \frac{1}{\theta}
[a,\hat{A}]+\frac{1}{\theta}[a^{\dag},\hat{\bar{A}}]+[\hat{\bar{A}},\hat{A}]-\frac{1}{k}[\hat{\Psi}^{\dag},
\hat{\Psi}]= \quad \quad \quad \quad \quad \quad \quad \quad \\
\quad \quad \quad \quad \quad \quad \quad
g^{-1}\left(\frac{1}{\theta}[a,\chi]-\frac{1}{\theta}[a^{\dag},
\chi^{\dag}]-2[\chi^{\dag},\chi] \right)g. \nonumber
 \end{eqnarray}
 Furthermore upon defining
 \begin{eqnarray}
\hat{\chi}\equiv-\frac{1}{\theta}\hat{h}^{-1} [a^{\dag},\hat{h}],
\quad \quad \hat{\chi}^{\dag}\equiv \frac{1}{\theta}\hat{h}^{-1}
[a,\hat{h}],
 \end{eqnarray}
for $\hat{h}$ in the gauge group this equation can be converted in
the nc chiral model equation
 \be
[a,\hat{h}^{-1}[a^{\dag},\hat{h}]]+[a^{\dag},\hat{h}^{-1}[a,\hat{h}]]=0.
\label{chiralnc} \ee In this sense given any solution $\hat{h}$ of
the nc chiral model, or alternatively any solution $\hat{\chi}$ of
(\ref{chi}), we could in principle obtain a solution of the nc
self-dual Chern-Simons equations (\ref{selfdualchsimonsop}). In
the ordinary commutative case there is a well established
procedure to construct the solutions of the chiral model equation
with have finite energy called the Uniton method \cite{uhlenbeck}.
In \cite{kimeon} was conjectured the extension of this method to
the nc plane and was explicitly constructed an specific solution
(the simplest) of the nc Liouville model ((\ref{kitoda}) taking
$N=2$). We would like to use this method to construct exact
solutions of our Toda field theories (\ref{Todaop}) and for this
reason  we will briefly discuss its details. The main idea is
based on one conjecture:{\it That the finite energy solutions
$\hat{h}$ of the nc $U(N)$ chiral model could be in principle
factorized uniquely as a product of uniton factors }\footnote{In
this section for notational simplicity we will write the
coordinates as $z,\bar{z}$ and the derivatives as usually, but
they should be understood as in (\ref{derivatives}). We keep the
hats in order to remember that we are working with operators. },
$\hat{h}=\hat{h}_0 \prod_{i=1}^{m}(2p_i-1)$, {\it where
a)}$\hat{h}_0$ {\it is a constant}, $m \leq N-1$, {\it b) each}
$p_i$ {\it is an hermitian} $N \times N$ {\it projection operator}
($p_i=p_i^{\dag}${\it and} $p_i=p_i^{2}$){\it c-)defining}
$\hat{h}_j=\hat{h}_0\prod_{i=1}^{j}(2p_i-1)$,{\it the following
linear relations must hold:}
\begin{eqnarray}
(1-p_i)\left(\partial + \frac{1}{2} \hat{h}_{i-1}^{-1} \partial
\hat{h}_{i-1} \right)p_i=0, \nonumber \\ (1-p_i)\left(
\hat{h}_{i-1}^{-1}\bar{\partial} \hat{h}_{i-1} \right)p_i=0
\label{conditions}.
\end{eqnarray}

This tell us that all the finite energy solutions of nc $U(2)$
chiral model can be written as $\hat{h}=\hat{h}_0(2p-1)$. That a
single uniton $\hat{h}=2p-1$, with $p$ the hermitian projection
operator satisfying the previous relations is a solution of the nc
chiral model equation (\ref{chiralnc}) is very simple to see. Then
the next step towards the construction of general solutions
involves the composition of uniton solutions. The holomorphic
projection operator can be written as the projection matrix \be
p=M(M^{\dag}M)^{-1}M^{\dag}, \ee where $M=M(z)$ is a rectangular
matrix that for $U(N)$ can be chosen as $N \times N^{\prime}$
matrix with $N^{\prime}< N$ \be M=\left(\begin{array}{cccc}
\hat{f}_{11}(z) & \hat{f}_{12}(z)& \dots & \hat{f}_{1N^{\prime}}(z) \\
\cdot & \cdot & \dots & \cdot \\
\hat{f}_{N1}(z) & \hat{f}_{N2}(z)& \dots &
\hat{f}_{NN^{\prime}}(z)
\end{array} \right),
\ee with $\hat{f}_{ij}(z)$ polynomials of $z$. Let us remark that
these projection operators are related to soliton solutions of the
$\mathbb{CP}^{N-1}$ model \cite{murunga} and in this case the
elements of $M$ must be polynomial for consistency reasons
\cite{kimeon}.

 At this point let us try to find the solutions of the Toda models
(\ref{Todaop}). So one start with the $N$ dimensional vector \be
u^{T}=(\hat{f}_1(z), \hat{f}_2(z), \dots , \hat{f}_N(z)), \ee and
then defines \be M_k=(\hat{u},
\partial \hat{u},
\partial^2 \hat{u}, \dots,
\partial^{k-1} \hat{u}), \ee
which is an $N\times k$ matrix. On the next step define the
projection operators \be p_k=M_k(M_k^{\dag}M_k)^{-1}M_k^{\dag}.
\ee  In this way, \be \hat{h}=(2p_1-1)(2p_2-1)\dots(2p_{N-1}-1)
\label{h} \ee is a solution of the nc $U(N)$ chiral model
equation. This claim is stated as a theorem in \cite{kimeon} and
proven there. The main idea is based on the fact that the vectors
$\hat{u},\partial \hat{u},\partial^2 \hat{u}, \dots,
\partial^{k-1} \hat{u}$ are considered as linear independent and from
these vectors through the Gram-Schmidt process \cite{murunga} unit
vectors $\hat{e}_i$ are constructed: \be
\hat{e}_i=(1-p_{i-1})\partial^{i-1}\hat{u}(\bar{\partial}^{i-1}\hat{u}^{\dag}(1-p_{i-1})\partial^{i-1}\hat{u})^{-\frac{1}{2}}
\quad \mathrm{for} \quad i=1,\dots ,N. \label{gramm} \ee Since
they span the same space as the vectors $\hat{u}, \partial \hat{
u},\partial^2 \hat{u}, \dots \partial^{k-1} \hat{u}$ then \be
p_k=\tilde{M}_k(\tilde{M}_k^{\dag}\tilde{M}_k)^{-1}\tilde{M}_k^{\dag},
\ee with $\tilde{M}_i=(\hat{e}_1,\hat{e}_2,\dots ,\hat{e}_i)$. And
since the vectors $\hat{e}_i$ with $i=1,\dots,k-1$ are orthonormal
($\hat{e}_i^{\dag}\hat{e}_j=\delta_{ij}$) it is possible to find a
simple expression for $p_i$: \be p_i=\sum_{j=1}^{i} \hat{e}_j
\hat{e}_j^{\dag}. \ee Notice however that these vectors depend on
both coordinates $z, \bar{z}$. The unitary matrix
$g=(\hat{e}_1,\hat{e}_2 \dots \hat{e}_N)$ diagonalizes each $p_i$
\be g^{-1}p_i g=\left(
\begin{array}{ccccccc} 1 & & & & &
& \\
& 1 & & & & & 0 \\
& & \ddots & & & & \\
& & & 1 & & & \\
& & & & 0 & & \\
& & & & & \ddots & \\
0 & & & & & & 0\\
\end{array}
\right) \label{diag}, \ee where the first $i$ entries on the
diagonal are $1$ and all the others are $0$. In the fo\-llo\-wing
we will verify that in fact (\ref{h}) satisfies the nc chiral
equation. This is not difficult to see since $p_i$ is a
holomorphic projection operator, i.e. $p_i\partial p_i=
\partial p_i$ for each $i=1 \dots N-1$, then
\begin{eqnarray}
\hat{h}^{-1}\partial
\hat{h}=\sum_{i=1}^{N-1}{(2p_{N-1}-1)\dots(2p_{i+1}-1)\partial
p_{i}(2p_{i+1}-1)\dots (2p_{N-1}-1)}.
\end{eqnarray}
By the other side from (\ref{diag}), \begin{eqnarray}
\partial(\hat{g}^{-1}p_i \hat{g})=0,
\end{eqnarray}
thus
\begin{eqnarray}
 \hat{g}^{-1}(\partial p_i)\hat{g}=[\hat{g}^{-1}\partial
\hat{g}, \hat{g}^{-1}p_i \hat{g}]. \end{eqnarray} From the
Gram-Schmidt procedure it is clear that $\partial e_i$ is a linear
combination of $e_1, \dots, e_{i+1}$ and by the orthonormality of
the basis $ e_{i}^{\dag}
\partial e_{j}=0$ for all $i > j+1$. Similarly since $\partial
(e_i^{\dag}e_j)=0$, then $e_i^{\dag}\partial e_j=-(e_j^{\dag}
\bar{\partial}e_i)^{\dag}= 0$ for all $j > i$, because
$\bar{\partial}e_i$ is a linear combination of $e_1, \dots,
e_{i}$. In this way the matrix $\hat{g}^{-1}\partial \hat{g}$ has
the simple form: \be \hat{g}^{-1}\partial \hat{g}=\left(
\begin{array}{ccccc}
\hat{e}_1^{\dag}\partial \hat{e}_1 & &  & & 0 \\
\hat{e}_2^{\dag}\partial \hat{e}_1 & \hat{e}_2^{\dag}\partial \hat{e}_2 & & & \\
 & \hat{e}_{i}^{\dag}\partial \hat{e}_{i-1} & \hat{e}_{i}^{\dag}\partial \hat{e}_i & & \\
 & \ddots & \ddots & & \\
 0 & & & \hat{e}_{N}^{\dag}\partial \hat{e}_{N-1} & \hat{e}_{N}^{\dag}\partial \hat{e}_N
 \end{array} \right). \label{gpg}
 \ee
This mean that $g^{-1}\partial p_i \hat{g} $ will have only one
element different from zero
 \be
( g^{-1}\partial p_i \hat{g})_{lm}=\hat{e}_{i+1}^{\dag}\partial
\hat{e}_{i}\delta_{l,i+1}\delta_{m,i}. \label{gg}
 \ee
It follows that $\hat{g}^{-1}[\partial p_i, p_j] \hat{g}=0$ for
$i<j$ and with this result
\begin{eqnarray}
\hat{h}^{-1}\partial \hat{h}=2\sum_{i=1}^{N-1}\partial p_{i} \quad
\mathrm{and} \quad \hat{h}^{-1}\bar{\partial}
\hat{h}=-2\sum_{i=1}^{N-1}\bar{\partial} p_{i},
\end{eqnarray}
since $\hat{h}$ is unitary and each $p_i$ is hermitian.
 Showing in
this way that the chiral equation is satisfied. Then as
$\hat{\chi}=\hat{g} \hat{\Psi} \hat{g}^{-1}$ and from (\ref{gpg}),
(\ref{gg}), (\ref{gaugepot}) the Toda solution takes the form
\begin{eqnarray}
\hat{\Psi}_{ij}=\hat{h}_i^{\dag}\delta_{i+1,j}=[\hat{g}^{-1}\hat{\chi}
\hat{g}]_{ij}=-\frac{1}{\theta}\delta_{i+1,j}e_{i+1}^{\dag}[a^{\dag},e_{i}],
\quad
i=1,\dots N-1, \nonumber \\
\hat{A}_{ij}=\hat{E}_i \delta_{i,j} = (\hat{\Psi}
-\frac{1}{\theta}\hat{g}^{-1}[a^{\dag},\hat{g}])\delta_{i,j}=-\frac{1}{\theta}\hat{e}_i^{\dag}[a^{\dag},\hat{e}_i]\delta_{i,j},
\quad i=1,\dots N. \label{solu0}
\end{eqnarray}
If we compare this expression with (\ref{relations1})  we can see
that
\begin{eqnarray}
\hat{g}^{-\frac{1}{2}}_{i}a^{\dag}\hat{g}^{\frac{1}{2}}_{i}&=&\hat{e}_i^{\dag}a^{\dag}\hat{e}_i,
\quad i=1\dots N \nonumber \\
\hat{g}_{i+1}^{\frac{1}{2}}\hat{g}_{i}^{-\frac{1}{2}}&=&-\frac{1}{\theta}\hat{e}^{\dag}_{i+1}[a^{\dag},\hat{e}_{i}],\quad
i=1\dots N-1. \label{solu}
\end{eqnarray}
 This algorithm allows to construct exact solutions of
 the nc Toda field theo\-ries (\ref{Todaop}). In the next section we will
 see how it explicitly works constructing some simple solutions of the nc
 Liouville model (\ref{eqnliouvillenc0}).

\subsection{NC Liouville exact solutions}

{\bf First solution:} Using the method explained in the previous
section, in the work \cite{kimeon} was calculated the simplest
exact solution to the nc Liouville model ((\ref{kitoda}) with
$N=2$). Here we will go one step forward and obtain one solution
of (\ref{eqnliouvillenc0}). Considering the vector $u^T=( z \quad
c )$, with $c$ a complex constant, the projection operator
$p=\hat{e}_1\hat{e}_1^{\dag}$ reads, \be p=\left(\begin{array}{cc}
             a \frac{1}{\mathbb{N}+|c|^2} a^{\dag}\quad & a
             \frac{1}{\mathbb{N}+|c|^2}\bar{c} \\
             c\frac{1}{\mathbb{N}+|c|^2}a^{\dag} \quad & |c|^2\frac{1}{\mathbb{N}+|c|^2}
             \end{array} \right).
\ee After computing the orthonormal vector $e_2$ the unitary
matrix $\hat{g}=(\hat{e}_1 \, \, \hat{e}_2)$ is expressed as
\footnote{Unitary in the sense that $gg^{\dag}=1$.}
 \be
 \hat{g}= \left(\begin{array}{cc}
             a\sqrt{\frac{1}{\mathbb{N}+|c|^2}}\quad  &
             \sqrt{\frac{|c|^2}{\mathbb{N}+\theta+c|^2}} \\
             c\sqrt{\frac{1}{\mathbb{N}+|c|^2}} \quad & \quad
             -a^{\dag}\frac{1}{\bar{c}}\sqrt{\frac{|c|^2}{\mathbb{N}+\theta+|c|^2}}
             \end{array} \right).
 \ee
Thus \be \hat{\Psi}=\hat{g}^{\dag}\partial p \hat{g}=
\left(\begin{array}{cc} 0 & 0
\\
\frac{\sqrt{|c|^2}}{\sqrt{\mathbb{N}+\theta+|c|^2}\sqrt{\mathbb{N}+|c|^2}}
& 0
\end{array} \right).
\ee Using that $a f(\mathbb{N})=f(\mathbb{N}+\theta)a$, the gauge
potentials $\hat{A}, \hat{\bar{A}}$ are expressed as
\begin{eqnarray} \hat{A}=-\frac{1}{\theta}\left(\begin{array}{cc}
  a^{\dag}\left(1-\frac{\sqrt{\mathbb{N}+|c|^2}}{\sqrt{\mathbb{N}+\theta+|c|^2}}\right)
   & 0 \\
   0 &
  a^{\dag}\left(1- \frac{\sqrt{\mathbb{N}+2\theta+|c|^2}}{\sqrt{\mathbb{N}+\theta+|c|^2}}\right)
   \end{array} \right).
\end{eqnarray}
Taking into account (\ref{psi1}) the simplest solution in terms of
$\hat{E}_1,\hat{E}_2,\hat{h}_1$ can be obtained from (\ref{solu0})
\begin{eqnarray}
\hat{E}_1=
\hat{g}_+^{-\frac{1}{2}}\left[a^{\dag},\hat{g}_+^{\frac{1}{2}}\right]
&=&
a^{\dag}\left(1-\frac{\sqrt{\mathbb{N}+|c|^2}}{\sqrt{\mathbb{N}+\theta+|c|^2}}\right),
\nonumber \\
\hat{E}_2=\hat{g}_-^{-\frac{1}{2}}\left[a^{\dag},
\hat{g}_-^{\frac{1}{2}}\right]
&=&a^{\dag}\left(1-\frac{\sqrt{\mathbb{N}+2+|c|^2}}{\sqrt{\mathbb{N}+\theta+|c|^2}}\right),
\nonumber \\
h_1^{\dag}=\hat{g}_-^{\frac{1}{2}}\hat{g}_+^{-\frac{1}{2}}&=&
\frac{\sqrt{|c|^2}}{\sqrt{\mathbb{N}+\theta+|c|^2}\sqrt{\mathbb{N}+|c|^2}}\,.
\label{solutliouville}
\end{eqnarray}
From the above two first equations it is obtained an exact
solution of our nc Liouville (\ref{eqnliouvillenc0})
\begin{eqnarray}
\hat{g}_+^{\frac{1}{2}}=\alpha(\sqrt{\mathbb{N}+|c|^2}), \quad
\hat{g}_-^{\frac{1}{2}}=\beta\left(\frac{1}{\sqrt{\mathbb{N}+\theta+|c|^2}}\right),
\end{eqnarray}
where $\alpha, \beta$ are two constants such that
$\alpha^{-1}\beta=\sqrt{|c|^2}$. The third condition in
(\ref{solutliouville}) fixes the relation among these constants.
We can choose $\alpha=1$ and $\beta=\sqrt{|c|^2}$ and the solution
is then written as
\begin{eqnarray}
\hat{g}_-=\frac{|c|^2}{\mathbb{N}+\theta+|c|^2}\, ,\quad
\hat{g}_+=\mathbb{N}+|c|^2\, .
\end{eqnarray}
In order to study the commutative limit of this solution we apply
the Weyl transform \cite{harvey} that leads to
\begin{eqnarray}
g_-=\frac{|c|^2}{z \star \bar{z}+|c|^2},\quad g_+=\bar{z}\star z
+|c|^2,
\end{eqnarray}
or equivalently to
\begin{eqnarray}
g_-=\frac{|c|^2}{\bar{z}z+\theta+|c|^2} \,,\quad
g_+=\bar{z}z-\theta+|c|^2 \,.
\end{eqnarray}
Considering a perturbative expansion in $\theta$,
\begin{eqnarray}
g_-&=&\frac{|c|^2}{\bar{z}z+|c|^2}\left(
1-\frac{\theta}{\bar{z}z+|c|^2} \right)+ O(\theta^2) \,, \nonumber
\\
g_+^{-1}&=&\frac{1}{\bar{z}z+|c|^2}\left(
1+\frac{\theta}{\bar{z}z+|c|^2} \right)+ O(\theta^2)\,,
\end{eqnarray}
it is not difficult to check that in fact up to first order in
$\theta$ this is a solution of the nc Liouville model \cite{me}
with $g_+=e_{\star}^{\phi_+}$ and $g_-=e_{\star}^{\phi_-}$ and
with $\varphi_1=\frac{1}{2}(\phi_+-\phi_-)$ and
$\varphi_0=\frac{1}{2}(\phi_++\phi_-)$. This solution in the
commutative limit $\theta \rightarrow 0$ reduces to the well known
Liouville solution:
 \be
\varphi_1=\ln \left( \frac{\bar{z}z+|c|^2}{|c|} \right). \ee

 {\bf Second solution:} As a second example let us now try to find
the next simplest solution fo\-llo\-wing the nc extension of the
uniton method of \cite{kimeon} outlined in the previous section.
For this purpose we will consider $u^T=(z^2 \, \, c )$, from where
it is computed the unit vector \be e_1=\left(
\begin{array}{c}
 a^2 \\
 c
\end{array} \right)\sqrt{\frac{1}{\mathbb{N}(\mathbb{N}-\theta)+
|c|^2}} \, , \ee and the projection operator \be
p=\left(\begin{array}{cc}
             a^2 \frac{1}{\mathbb{N}(\mathbb{N}-\theta)+|c|^2} {a^{\dag}}^2 &
             a^2
             \frac{1}{\mathbb{N}(\mathbb{N}-\theta)+|c|^2}\bar{c} \\
             c\frac{1}{\mathbb{N}(\mathbb{N}-\theta)+|c|^2}{a^{\dag}}^2 & |c|^2\frac{1}{\mathbb{N}(\mathbb{N}-\theta)+|c|^2}
             \end{array} \right).
\ee
 On the next step we
compute the orthonormal vector $e_2$ using the expression
(\ref{gramm}) and it reads  \be e_2=\left(
\begin{array}{c}
 a \\
-\frac{c}{|c|^2}a^{\dag}\mathbb{N}
\end{array} \right)\sqrt{\frac{1}{\mathbb{N}}}\sqrt{\frac{|c|^2}{(\mathbb{N}+\theta)\mathbb{N}+
|c|^2}} \,. \ee By means of the expressions (\ref{solu}) it is
computed another solution of the nc Liouville model,
\begin{eqnarray}
g_+^{-\frac{1}{2}}a^{\dag}g_+^{\frac{1}{2}}=e_1^{\dag}a^{\dag}e_1&=&a^{\dag}\sqrt{\frac{\mathbb{N}(\mathbb{N}-\theta)+|c|^2}{(\mathbb{N}+\theta)\mathbb{N}+|c|^2}}
\, ,
\nonumber \\
g_-^{-\frac{1}{2}}a^{\dag}g_-^{\frac{1}{2}}=e_2^{\dag}a^{\dag}e_2&=&a^{\dag}\sqrt{\frac{\mathbb{N}}{(\mathbb{N}+\theta)}}\sqrt{\frac{(\mathbb{N}+2\theta)(\mathbb{N}+\theta)+|c|^2}{(\mathbb{N}+\theta)\mathbb{N}+|c|^2}}\,.
\end{eqnarray}
From where we get that
\begin{eqnarray}
g_+^{\frac{1}{2}}=\alpha
\sqrt{\mathbb{N}(\mathbb{N}-\theta)+|c|^2} \,, \quad
g_-^{\frac{1}{2}}=\beta
\sqrt{\frac{\mathbb{N}}{(\mathbb{N}+\theta)\mathbb{N}+|c|^2}} \,.
\end{eqnarray}
The constants are related through the condition \be
h_1^{\dag}=-\frac{1}{\theta}e_2^{\dag}[a^{\dag},e_1]=g_-^{\frac{1}{2}}g_+^{-\frac{1}{2}}=\sqrt{\frac{\mathbb{N}}{(\mathbb{N}+\theta)\mathbb{N}+|c|^2}}
\sqrt{\frac{4|c|^2}{\mathbb{N}(\mathbb{N}-\theta)+|c|^2}} \,, \ee
from where we obtain that \be \alpha^{-1}\beta= 2\sqrt{|c|^2}.\ee
Choosing $\alpha=1$ and $\beta=2\sqrt{|c|^2}$, one solution is
then
\begin{eqnarray}
g_-=\frac{4|c|^2\mathbb{N}}{(\mathbb{N}+\theta)\mathbb{N}+|c|^2}
\, ,\quad g_+=\mathbb{N}(\mathbb{N}-\theta)+|c|^2 \,,
\end{eqnarray}
that in the commutative limit leads to \be \varphi_1=\ln \left(
\frac{(\bar{z}z)^2+|c|^2}{2|c|\sqrt{\bar{z}z}}\right), \ee another
known Liouville solution.

 {\bf Third solution:} A more general solution that include
the above solutions as particular examples could be computed using
the uniton method. Take now the vector as $u^T=( z^m \, c)$.
  The matrix projector $p$ is in this case expressed as
 \be p=\left(\begin{array}{cc}
             a^m \frac{1}{\mathbb{N}_m+|c|^2} {a^{\dag}}^m &
             a^m
             \frac{1}{\mathbb{N}_m+|c|^2}\bar{c} \\
             c\frac{1}{\mathbb{N}_m+|c|^2}{a^{\dag}}^m & |c|^2\frac{1}{\mathbb{N}_m+|c|^2}
             \end{array} \right),
\ee
 where
$\mathbb{N}_m=\mathbb{N}(\mathbb{N}-\theta)\dots(\mathbb{N}-m\theta+\theta)$.
The matrix $g$ is equal to
 \be
 g= \left(\begin{array}{cc}
             a^m\sqrt{\frac{1}{\mathbb{N}_m+|c|^2}}\quad  &
             a^{m-1}\sqrt{\frac{|c|^2}{(\mathbb{N}+\theta)_m+|c|^2}}\frac{1}{\sqrt{\mathbb{N}_{m-1}}} \\
             c\sqrt{\frac{1}{\mathbb{N}_m+|c|^2}} \quad & \quad
             -a^{\dag}\frac{1}{\bar{c}}\sqrt{\frac{|c|^2\mathbb{N}_{m-1}}{(\mathbb{N}+\theta)_m+|c|^2}}
             \end{array} \right),
 \ee
where $\mathbb{N}_{m-1}=\mathbb{N}(\mathbb{N}-\theta)\dots
(\mathbb{N}-(m-2)\theta)$ and
$(\mathbb{N}+1)_{m}=(\mathbb{N}+\theta)\mathbb{N}(\mathbb{N}-\theta)\dots
(\mathbb{N}-(m-2)\theta)$. Taking into account (\ref{solu}) it is
obtained
\begin{eqnarray}
g_+^{-\frac{1}{2}}a^{\dag}g_+^{\frac{1}{2}}=e_1^{\dag}a^{\dag}e_1&=&a^{\dag}\sqrt{\frac{\mathbb{N}_m+|c|^2}{(\mathbb{N}+\theta)_{m}+|c|^2}}
\, ,
\nonumber \\
g_-^{-\frac{1}{2}}a^{\dag}g_-^{\frac{1}{2}}=e_2^{\dag}a^{\dag}e_2&=&a^{\dag}\sqrt{\frac{\mathbb{N}_{m-1}}{(\mathbb{N}+\theta)_{m-1}}}\sqrt{\frac{(\mathbb{N}+2\theta)_m+|c|^2}{(\mathbb{N}+\theta)_{m}+|c|^2}}
\, ,
\end{eqnarray}
from where we get that,
\begin{eqnarray}
g_+^{\frac{1}{2}}=\alpha \sqrt{\mathbb{N}_m+|c|^2} \, , \quad
g_-^{\frac{1}{2}}=\beta
\sqrt{\frac{\mathbb{N}_{m-1}}{(\mathbb{N}+\theta)_m+|c|^2}} \,.
\end{eqnarray}
Once again the constants are related through the condition \be
h_1^{\dag}=-\frac{1}{\theta}e_2^{\dag}[a^{\dag},e_1]=g_-^{\frac{1}{2}}g_+^{-\frac{1}{2}}=\sqrt{\frac{\mathbb{N}_{m-1}}{(\mathbb{N}+\theta)_m+|c|^2}}
\sqrt{\frac{m^2|c| ^2}{\mathbb{N}_m+|c|^2}} \, , \ee from where we
get that \be \alpha^{-1}\beta=m \sqrt{|c|^2}.\ee Choosing
$\alpha=1$ and $\beta=m\sqrt{|c|^2}$, the solution is then
\begin{eqnarray}
g_-=\frac{m^2|c|^2\mathbb{N}_{m-1}}{(\mathbb{N}+\theta)_m+|c|^2},\quad
g_+=\mathbb{N}_m+|c|^2.
\end{eqnarray}
This solution in the commutative limit reduces to another
classical Liouville solution:
\begin{eqnarray}
\varphi_1=\ln\left(\frac{(\bar{z}z)^m+|c|^2}{m|c|\sqrt{(\bar{z}z)^{m-1}}}
\right).
\end{eqnarray}
In this section we have seen how although the nc extension of the
uniton method is still not proven \cite{kimeon}, it can be used to
compute exact solutions of the nc Toda models (\ref{Todaop}).
Particularly we have constructed exact solutions of the nc
Liouville model (\ref{eqnliouvillenc0}) and these solutions reduce
in the commutative limit to known solutions of the ordinary
Liouville model, what in a certain sense gives validity to the
method. The construction of exact solutions of other Toda models
($N>3$) is straightforward, although with much more complicated
calculations involved.


\section{Conclusions}

In this paper we have studied in a more detailed way the relation
between the nc self-dual Chern-Simons system and the nc
Leznov-Saveliev equations. We have seen how from the NCSDCS system
it is possible to define the Toda field theories as systems of
second order differential equations and still it is possible to
construct exact solutions using the nc extension of the uniton
method proposed in \cite{kimeon}. The solutions explicitly
constructed for the nc Liou\-ville model lead to known solutions
in the commutative limit, what in a certain way validate the
method. Since the NCSDCS system can be obtained from the nc
self-dual Yang-Mills equations in four dimensions through a
dimensional reduction process \cite{me}, the nc Toda field
theories could possible have a physical picture inside D-branes
systems. Finally we could say that although the complete
integrability properties of these theories remains to be
investigated, the nc Toda field theories constructed in \cite{me}
posses in fact some integrable-like properties: an infinite number
of conserved charges, \footnote{Some of these charges can be
constructed using the bicomplex formalism (see \cite{dimakis,
me}).} exact solutions, and they are  reductions of the nc
self-dual Yang-Mills equations in four dimensions that in
\cite{takasaki} was shown to be classically integrable.
\section*{Acknowledgments}

 I thank to the Physics Department-UnB for the kind hospitality and specially to Prof. Ademir Santana
 for the invitation to give a seminar where part of this work was presented. This work has been
 supported by CNPq-FAPEMAT.

\end{document}